# Room Temperature Nanoscale Ferroelectricity in Magnetoelectric GaFeO$_3$ Epitaxial Thin Films

*Somdutta Mukherjee$^{\S 1}$, Amritendu Roy$^{\S 2}$, Sushil Auluck$^3$, Rajendra Prasad$^1$, Rajeev Gupta$^{1,4}$ and Ashish Garg$^2$*

$^1$Department of Physics, Indian Institute of Technology Kanpur, Kanpur- 208016, India
$^2$Department of Materials Science and Engineering, Indian Institute of Technology Kanpur, Kanpur- 208016, India
$^3$National Physical Laboratory, K.S Krishnan Marg, New Delhi 110012, India
$^4$Materials Science Programme, Indian Institute of Technology Kanpur, Kanpur- 208016, India

**Abstract**

We demonstrate room temperature ferroelectricity in the epitaxial thin films of magnetoelectric GaFeO$_3$. Piezo-force measurements show a 180$^o$ phase shift of piezoresponse upon switching the electric field indicating nanoscale ferroelectricity in epitaxial thin films of gallium ferrite. Further, temperature dependent impedance analysis with and without the presence of an external magnetic field clearly reveals a pronounced magneto-dielectric effect across the magnetic transition temperature. In addition, our first principles calculations show that Fe ions are not only responsible for ferrimagnetism as observed earlier, but also give rise to the observed ferroelectricity, making GFO an unique single phase multiferroic.

---

Pursuit of multifunctionalities in single phase or composite materials has led to sustained research on multiferroic materials. These materials, mostly artificially synthesized, can give rise to a variety of novel applications such as spintronic and data storage devices, sensors and actuators. [1, 2] Rare occurrence of natural multiferroic materials has led to extensive search for materials systems [3, 4] and over the last decade, a combination of advanced synthesis and characterization techniques [5, 6] and state-of-the-art first-principles studies [7, 8] have predicted numerous multiferroic materials. However, with the exception of ferroelectric-antiferromagnetic $BiFeO_3$, most materials demonstrate multiferroism at very low temperatures.[5, 9] Thus, it is vital to explore new multiferroic materials demonstrating multiferroic effect with significant magnetoelectric (ME) coupling near or above room temperature (RT) in order to realize their technological promise.

Gallium ferrite ($Ga_{2-x}Fe_xO_3$ or GFO) is a room temperature piezoelectric[10-14] and near room temperature ferrimagnetic material with its magnetic transition temperature tunable to room temperature and above by tailoring its composition.[15] Though, the magnetic characteristics of GFO are widely studied,[10, 13, 16-18] intriguingly there is no evidence of its ferroelectric nature. While an early report[19] attributed asymmetrically placed Ga1 ions within the unit cell responsible for observed piezoelectric response of GFO, recent first-principles calculations[20] showed that within the inherently distorted structure of GFO, large ionic displacements with respect to the centrosymmetric positions result in a large spontaneous polarization in the ground state[20] and even hint towards possible ferroelectric switching.[21] Thus, inability to observe saturated ferroelectric hysteresis loops (if any) in GFO bulk and single crystal samples is likely to emanate from the measurement difficulties, possibly due to substantial electrical leakage above 200 K.[22-24] On the other hand, epitaxial thin films of pure and doped GFO, grown on a variety of single crystalline substrates show a large reduction in the leakage current[24, 25] and are more likely to demonstrate ferroelectric behavior if probed locally.

In this work, we report RT nanoscale ferroelectric switching in (010)-oriented epitaxial thin films of GFO, along with the presence of near RT ferrimagnetism. Subsequent first-principles calculations reveal that Fe ions are responsible for both ferroelectricity and ferrimagnetism making GFO an unique multiferroic material.[3] In the remaining paragraphs, we first describe the structural analysis of as grown thin films followed by their electrical and magnetic characterization and first-principles calculations results substantiating ferroelectricity as well as magnetoelectric coupling.

GaFeO$_3$ (GFO) thin films were grown on commercially available single crystalline cubic yittria stabilized zirconia, YSZ (001) substrate (lattice parameter, $a_{YSZ}$ = 5.125 Å). For electrical characterization, transparent conducting indium tin oxide (ITO) was used as bottom electrode. Both GFO and ITO were grown using pulsed laser deposition (PLD) technique with KrF excimer laser ($\lambda$ = 248 nm) operated at 3 Hz and 10 Hz, respectively. GFO films of 200 nm thickness were grown at 800 °C in an oxygen ambient ($p_{O2}$ ~ 0.53 mbar) using a laser fluence of 2 J cm$^{-2}$ from a stoichiometric target of gallium ferrite[15] while ITO films of 40 nm thickness were grown using a laser fluence of 1 J cm$^{-2}$ at 600°C at $p_{O2}$ ~ 1×10$^{-4}$ mbar using an ITO target. The films were subsequently cooled at 1°C min$^{-1}$ to 300° C at the same O$_2$ pressure used for GFO deposition followed by natural cooling to room temperature. X-ray diffraction of the as grown film was performed using PANalytical X'Pert Pro MRD diffractometer using CuK$\alpha$ radiation. Surface topography and domain structure were studied using scanning probe microscope (Asylum Research) equipped with Olympus AC240TS Ti/Ir tip operated at resonance frequency. The same setup was used to carry out switching spectroscopy mapping (SSPFM) measurements with Rocky mountain cantilever equipped with 25Pt400B solid pt probe. For SSPFM measurement, we used Dual Ac Resonance Tracking (DART) mode. For impedance measurement, Pt top electrode (~ 200 μm diameter) was deposited by sputtering, using shadow mask technique. Impedance data was acquired using Agilent Impedance analyzer 4294A connected to a commercial ARS He close cycle cryo-probe station placed between two magnetic pole pieces.

First-principles calculations were performed using density functional theory within the generalized gradient approximation (GGA+U) with Perdew and Wang (PW91) functional[26] as implemented in Vienna *ab-initio* simulation package (VASP) [27] and using rotationally invariant approach [28] with onsite Coulomb potential $U_{eff}$ = 5.5 eV to treat the localized *d* electrons of Fe ions. This value of $U_{eff}$ has been found to yield reasonable agreement between calculated and experimental magnetic moments of Fe ions in GFO. Further, small variation of the value of $U_{eff}$ was found not to alter the structural stability. We verified the consistency of our calculations by repeating the calculations using GGA method with the optimized version of Perdew-Burke-Ernzerhof functional for solids (PBEsol).[29] The GGA functionals PW91 and PBEsol also yielded similar results. More information on calculation details can be found elsewhere.[20]

**Figure. 1** (a) shows the θ-2θ X-Ray diffraction (XRD) pattern of phase pure and 200 nm thin GFO films deposited on (001)-oriented yittria stabilized zirconia (YSZ) substrates buffered with a 40 nm indium tin oxide (ITO) layer, also acting as the bottom electrode. The figure shows only {010} type of peaks of GFO (orthorhombic *Pc2₁n* symmetry) along with (001) peaks of ITO and YSZ indicating an out of plane epitaxial relationship as $(010)_{GFO}$ ∥ $(001)_{ITO}$ ∥ $(001)_{YSZ}$. Calculated out of plane lattice parameter, *b* ~ 9.4012 Å, is in excellent agreement (~ 0.02% difference) with *b*-axis lattice parameter of bulk single crystal[10] indicating that the film is fully relaxed along film's *b*-axis. A small lattice mismatch between ITO ($a_{ITO}$ ~ 1.016 nm) and diagonal $[(a_{YSZ}^2 + c_{YSZ}^2)^{1/2}]$ of in-plane lattice parameters of GFO of 0.4% [30] and lattice mismatch between $a_{ITO}$ and $2a_{YSZ}$ of 1.13% indicates that GFO film is coherently strained within the substrate plane, also demonstrated by the corresponding reciprocal space map (Figure 1(c)). Nature of in-plane orientation of the film was determined by performing a ϕ-scan corresponding to (221) peak of GFO, (222) peak of ITO electrode and (111) peak of the YSZ substrate. As shown in Figure. 1(b), the presence of four equally spaced peaks for ITO and YSZ indicates that ITO films maintain similar crystallographic orientation as of YSZ. However we observe 12 peaks in the ϕ-scan of GFO films indicating existence of different growth variants. Different growth variants are commonly seen in epitaxial thin films of oxides [31, 32] which are largely due to tendency of single crystal oxide substrates to cleave along certain crystallographic planes leaving facets on the substrate surface.

Topography of a 200 nm thick GFO film estimates average grain size ~ 96 nm and RMS roughness ~ 9.5 nm. Converse piezoelectric effect with lock-in technique was employed to study the local piezoelectric switching behavior and to estimate the $d_{33}$ coefficient. PFM was used in spectroscopic mode where measurement was done in a fixed tip position with a dc bias voltage swept in a cyclic manner. The dependence of local piezoelectric vibration on the corresponding voltage sweep is referred as local piezoelectric hysteresis loop. On a macroscopic scale, there will be weak field dependence of piezoelectric coefficient, $d_{33}$, with continuously varying bias field. To verify the presence of ferroelectricity, we applied a sequence of dc voltage in a triangular saw-tooth form in an attempt to switch the polarization with a 2 V ac voltage simultaneously applied in order to record the corresponding piezoresponse. To minimize the effect of electrostatic interaction, piezoresponse was measured during "off" state at each step, and phase voltage hysteresis loop is evident. The $d_{33}$ dependence of the polarization can be obtained by local bias voltage switching.

We investigated the piezo- and ferroelectric behavior of these films using piezoresponse force microscopy (PFM). Figure 2 (a) and (b) show PFM amplitude and phase images acquired over 1.25×1.25 μm² area in PFM Dual AC Resonance Tracking imaging mode, using a cantilever of stiffness 2 N m$^{-1}$ and a Ti/Ir tip. Figure 2(a) shows the out-of-plane polarization as depicted by the bright yellow regions while Figure 2(b) shows the presence of antiparallel nanodomains with concurrently minor presence of domains with intermediate domain angle. For studying local piezoelectric and ferroelectric switching, we also plotted the phase and butterfly amplitude loops upon sweeping bias voltage. Figure 2 (c) and (d) show the corresponding amplitude (A) and phase (ϕ) loops as a function of dc bias voltage. The butterfly loop in Figure 2(c) reveals the first harmonic signal under applied dc bias field and is signature of piezoelectric response of the thin films. The piezoresponse tends to saturate at relatively high voltages suggesting that the response is piezoelectric instead of electrostatic. The phase (ϕ) corresponds to the phase of piezoresponse and its reversal with voltage is shown in Figure 2(d). This reversal occurs beyond a coercive voltage, ~ -2.9 V at negative side and ~ 3.6 V at positive side while the phase contrast is ~180° clearly suggesting polarization switching and thus, ferroelectric character of our GFO thin films.

Having shown RT ferroelectricity, it would be interesting to explore possible magnetoelectric interaction in GFO thin films since such an effect would increase the material's acceptability as a close to room temperature multiferroic memory material. We then probed possible magnetoelectric coupling by performing temperature dependent impedance spectroscopic analysis, from 50 K to 325 K. Figure. 3 presents the plot of real part of dielectric constant ($\varepsilon'$) at frequencies 1, 5, 10, 25, 50 and 100 kHz. The Figure shows that the onset of increase in the dielectric constant is approximately at 150 K at 1 kHz, shifting to higher temperatures at higher frequencies. However, plots show a hump in the dielectric constant ($\varepsilon'$) at ~235 K, in the vicinity of ferri to paramagnetic transition temperature (as shown in the bottom inset). Such deviation in the dielectric constant from a typical temperature dependent dielectric behavior is considered as an indication of the magnetoelectric coupling in GFO. The temperature ($T_m$) corresponding to peak position in $\varepsilon'$ exhibits a weak frequency dependence and shifts towards higher temperature from 230 K at 1 kHz to 240 K at 100 kHz. Further, we measured the dielectric constant at 10 kHz in presence of two different magnetic fields ($\mu_0 H$ = 0.25 and 0.5 T) across $T_m$. As shown in the top inset of Figure 3, with increasing magnetic field, the dielectric anomaly across $T_m$ becomes suppressed, providing unambiguous evidence of magnetoelectric coupling in GFO thin films. The calculated magnetodielectric coefficient ($\varepsilon(H)$- $\varepsilon(0)$/ $\varepsilon(0)$) at 0.5 Tesla is -0.154. Interestingly, this value is nearly one order of magnitude higher than those observed in polycrystalline GFO [22]. This increase in the coupling strength of epitaxial GFO films could arise due to several reasons: epitaxial strain, constrained 2-D film geometry, or microstructure and it would be of further interest to probe the exact cause such as by carrying out thickness dependent studies.

To understand the mechanism of nanoscale ferroelectricity in epitaxial gallium ferrite thin films, we further performed first-principles calculations on the ground state structure of GFO using GGA+U technique. Initially, we identified orthorhombic *Pnna* as the possible centrosymmetric structure of GFO which transforms to noncentrosymmetric *Pc2$_1$n* (*Pna2$_1$*, according to international table of crystallography) structure, using the calculation approaches reported earlier[20, 21]. Using optimized structures of centrosymmetric *Pnna* and noncentrosymmetric *Pna2$_1$* phase of GFO (say P↑), we constructed a second *Pna2$_1$* cell which is a mirror image of optimized *Pna2$_1$* (P↓) structure across the displacement coordinate with respect to the centrosymmetric *Pnna* cell. The calculations show that the two polarization states have identical ground state energies, a key signature of ferroelectricity in a material. A comparison between the centrosymmetric and polar structures, as shown in Figure. 4 (a), shows that there is a large displacement of Fe atoms with respect to other atoms with particularly large distortion seen for Fe2-O octahedra when GFO undergoes transformation to a noncentrosymmetric structure. Our calculations estimate that both the Fe ions in the *Pna2$_1$* structure displace by a much larger distance along the polar direction ($|u| \sim 0.22$Å) in comparison to the Ga ions ($|u| \sim 0.13$Å) upon *Pnna*→*Pna2$_1$* transformation. Such a large displacement of atoms is expected to cost a large energy and could possibly hint why a thermally induced phase transition in GFO has been elusive. Based on these displacements, the calculated spontaneous polarization of the polar structure is 0.28 C.m$^{-2}$ using the nominal ionic charges of the constituent ions and 0.33 C m$^{-2}$ using Born effective charges which are in close agreement with other reports.[21] Our calculations also show that the polarization contribution from the Fe ions is significantly larger than that by the Ga ions and therefore suggest that ferroelectricity in GFO is brought about predominantly via displacement of Fe ions.

The calculated energy difference between centrosymmetric and noncentrosymmertic structures is 0.61 eV f.u$^{-1}$ for GFO using GGA+U and is in agreement with literature.[21] However, the magnitude of the energy barrier is quite large in comparison to common perovskite ferroelectric oxides such as $PbTiO_3$ and $PbZrO_3$[30]. The abnormally large change in the energy upon ferroelectric phase transition cannot be explained by the large structural distortion alone and lack of any structural phase transition makes it even more puzzling. Several temperature dependent experimental studies [10, 33, 34] do not show any phase transition from non-centrosymmetric to centrosymmetric structure at least until 1368 K implying that its ferroelectric $T_c$ is even higher. As a consequence, the energy difference between two structures of GFO and the accompanying distortion should only be considered qualitatively. In this context, our observations of saturated loops in epitaxially strained GFO thin film samples are suggestive of a reduced energy barrier between centrosymmetric and noncentrosymmetric structures.[35] An alternative explanation for the observed discrepancy between the calculated energy barrier and observed ferroelectric switching at room temperature in GFO films could be the presence of domains in these samples as domains in ferroelectrics are known to significantly reduce the energy barrier required for switching.[36, 37] Further, for sustainable ferroelectric polarization, in addition to showing a double potential well, GFO must remain insulating all along during ferroelecrric switching i.e. from P↑ to P↓. Spin-resolved total density of states calculations at every point on the switching path, as shown in the insets of Figure. 4(b), demonstrate insulating nature of the system during polarization switching.

As far as mechanism of multiferroism in GFO is concerned, we now combine the reasons of observed ferroelectricity and magnetism together to evolve a collective picture. Previous theoretical and experimental studies[10, 13, 38] have conclusively shown that the observed ferrimagnetism in GFO is due to cationic site disorder where some Fe ions occupy Ga sites. In addition, as shown in the preceding paragraphs, ferroelectricity also emanates from the displacement of Fe ions from the centrosymmetric structure along *c*-axis of GFO (*b*-axis for conventional $Pc2_1n$ symmetry). These observations together suggest that the multiferroism in GFO originates from the same ionic species *i.e.* Fe ions, making it a unique multiferroic. Such mechanism of multiferroism is in contrast to the conventional perception that ferroelectricity (empty cation *d*-shell) and magnetism (partially filled cation *d*-shell) exclude each other.[3]

Having shown that the same ion is responsible for magnetism and ferroelectricity in in GFO, we now explore the magnetoelectric coupling in GFO (experimental evidence shown in Fig. 3) by calculating the energy difference between ferroelectric and paraelectric phases upon changing the magnetic spin configuration. We calculated the energy barrier ($\Delta E$) between ferroelectric and paraelectric phases of GFO coexisting with different spin structures, *viz.* antiferromagentic spin ordering and unpolarized spins (non-magnetic). The calculations show that the energy barrier is lower by 60 meV for an antiferromagnetic spin configuration, also bolstering the fact that the antiferromagentic spin structure of ferroelectric phase of GFO is more stable. This, in conjunction with previous observations of presence of magneto-structural coupling[33, 38] in GFO, shows that ferroelectric GFO possesses both magnetoelectric and magnetostructural coupling. Overall, presence of ferroelectricity, ferrimagnetism, magneto-electric-structural coupling in GFO thin films in the vicinity of room temperature make GFO an exciting material from the perspective multi-mode devices such as sensors and memories.

In summary, we have shown a first conclusive experimental evidence of nanoscale room temperature ferroelectricity in epitaxial thin films of gallium ferrite along with presence of magnetoelectric coupling. Interestingly, our first-principles calculations suggest that it is the Fe ions which are responsible for both ferroelectricity as well as ferrimagnetism. This finding is crucial as it establishes GFO as a near room temperature multiferroic and as a single phase material showing both ferroelectric and ferrimagnetic ordering, obviating the need of exchange bias multilayer junctions.

The work was partially funded by Department of Science and technology, Govt. of India through the project SR/S2/CMP-0098/2010. Authors thank Amir Moshar (Asylum Research) for PFM measurements, Anurag Gupta (NPL, New Delhi, India) for magnetic measurements and DST Nanoscience unit for XRD studies. Authors thank Dr. D. Stoeffler (IPCMS, Strasbourg) and Prof.

J.F. Scott (Cambridge University) for fruitful discussions and SA thanks NPL, New Delhi for financial assistance.10

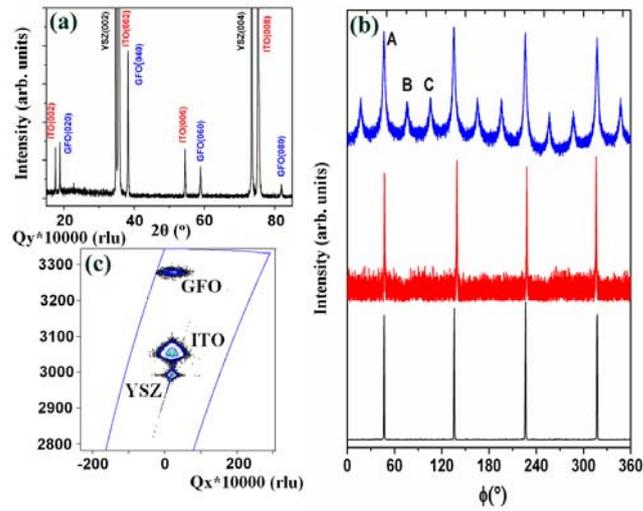

**Figure 1.** (a) θ-2θ XRD scan showing (010) and (001) orientations of GFO and ITO layers deposited on YSZ (001) substrate. (b) XRD ϕ-scan of {111} planes of YSZ (bottom), ITO (middle) and {221} planes of GFO (top) exhibiting four-fold symmetry for YSZ and ITO conducting layer while GFO showing three variant epitaxy. (c) Reciprocal space map (RSM) for 200 nm GFO film on ITO buffered YSZ substrate near the (020) reflection of the orthorhombic phase.

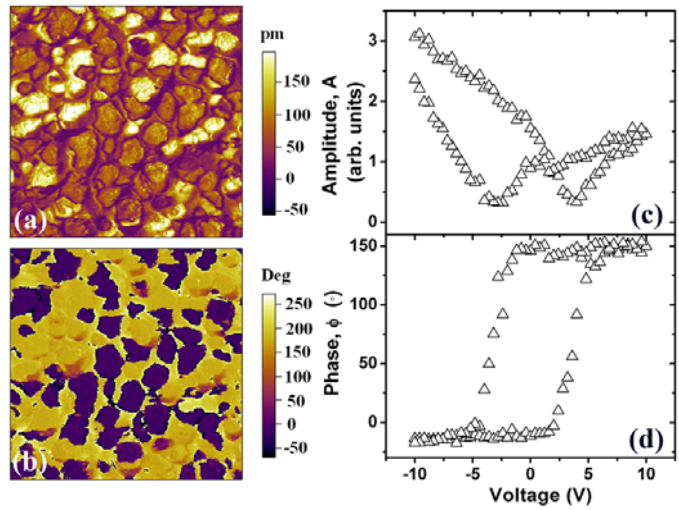

**Figure 2.** (a) Out of the plane PFM amplitude and (b) PFM phase micrographs of GFO (200 nm)/ITO (40 nm)/YSZ showing mosaic domain structure. Local piezoelectric response amplitude (c) and phase (d) on b-axis oriented gallium ferrite thin film measured using switching spectroscopy (SS) PFM mode.

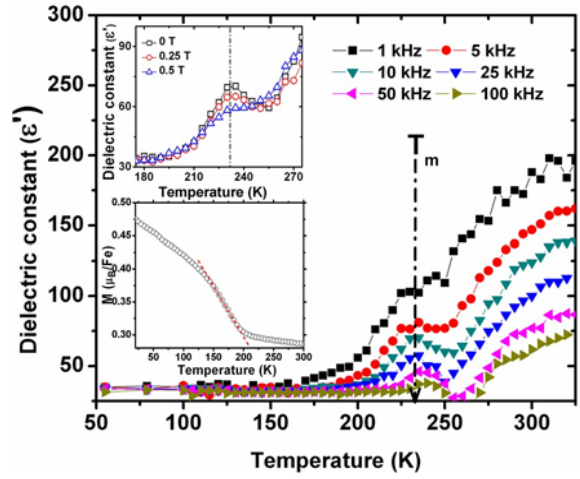

**Figure 3.** Real part of dielectric constant (ε′) vs. temperature plots measured at different frequencies showing a dielectric anomaly at ~ 235 K, close to ferri to paramagnetic transition temperature ($T_c$). Dielectric anomaly temperature ($T_m$) is marked by a dash-dot line. Top inset showing ε′ vs. temperature plot measured at 10 kHz in presence of different magnetic fields. It is observed that with increasing magnetic field the dielectric anomaly vanishes. Bottom inset plots magnetization as a function of temperature clearly showing the magnetic transition temperature ($T_c$).

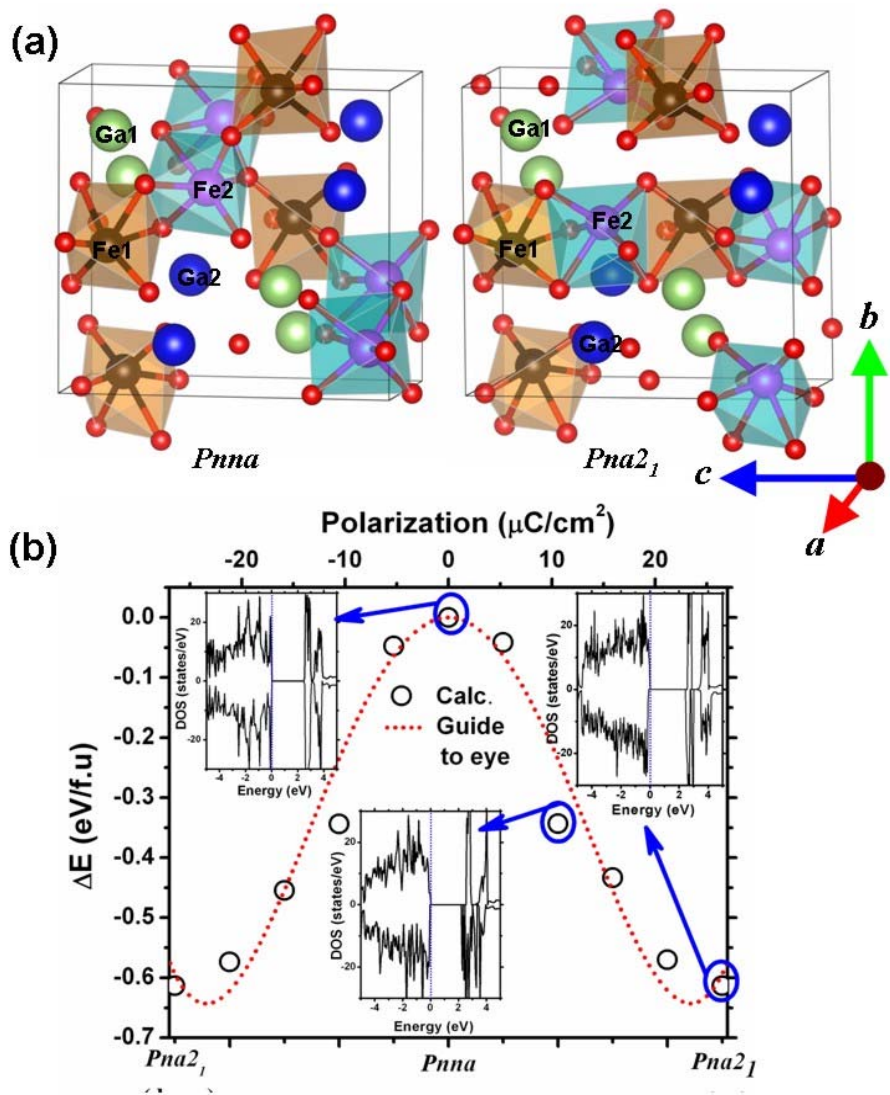

**Figure 4.** (a) Structural models of centrosymmetric (*Pnna*) and noncentrosymmetroic polar structures (*Pna2₁*) depicting the relative changes in the ionic positions, particularly for Fe-O octahedra, upon structural transformation (red spheres depict O atoms) (b) Switching path between two polar states via centrosymmetric phase. Insets show spin-resolved total density of states at different points on the transition path.